\documentclass[10pt,twocolumn]{article}
\usepackage[a4paper,left=0.75in, right=0.75in, vmargin=0.62in]{geometry}
\usepackage{amsmath,amssymb,amsfonts}
\usepackage[T1]{fontenc}
\usepackage{lmodern}
\usepackage{physics}
\usepackage{graphicx}
\usepackage{caption}
\usepackage{subcaption}
\usepackage{float}
\usepackage{cite}
\usepackage{authblk}
\usepackage{setspace}
\usepackage{times}
\usepackage{booktabs}
\usepackage{multirow}
\usepackage{abstract}
\usepackage{hyperref}
\usepackage{stfloats}
\usepackage{placeins}
\usepackage{balance}
\usepackage{parskip}

\title{\bfseries
Benchmarking Quantum Algorithmic Resilience for CVaR Portfolio Optimization: The Expressibility-Coherence Trade-off
}

\author[1]{Prashik N. Somkuwar}
\author[2]{K. Srinivasan}
\author[3]{G. Raghavan}

\affil[1, 2, 3]{School of Quantum Technology, Defence Institute of Advanced Technology, Pune, India\\
\vspace{0.15cm}
$^{1}$\texttt{prashik{\_}s@mt.iitr.ac.in} \quad 
$^{2}$ \texttt{srinik@diat.ac.in}  \quad
$^{3}$\texttt{graghavan@diat.ac.in}}

\date{}
\vspace{-0.4cm}

\begin{document}
\twocolumn[
\maketitle

\begin{abstract}
\vspace{0.4cm}
Quantum combinatorial optimization offers theoretical advantages for complex financial modeling, but physical implementation on Noisy Intermediate Scale Quantum (NISQ) devices is severely constrained by hardware topology. This study presents a hardware benchmarking analysis between a Hardware Efficient Variational Quantum Neural Network (HE-VQNN) and the Warm Start Quantum Approximate Optimization Algorithm (WS-QAOA) for a hybrid Mean Variance and Conditional Value at Risk (CVaR) portfolio objective. By implementing a novel classical quantum hybrid proxy matrix to bypass the CVaR auxiliary qubit bottleneck, we map up to 16 assets from the NIFTY 50 index onto an IBM heavy hex processor. We systematically quantify algorithmic resilience to the "SWAP tax" incurred during routing. Empirical results reveal a critical operational trade-off: WS-QAOA provides exact theoretical mapping but suffers catastrophic hardware decoherence due to exponential nonlocal gate overhead. Conversely, HE-VQNN preserves hardware coherence but lacks the mathematical expressibility to capture dense tail risk asset correlations. This study exposes the limitations of dense financial optimization on current architectures forces an nonviable choice between algorithmic inexpressibility and hardware decoherence. This is indicative of a deeper limitation as to what can and cannot be done with NISQ computers lacking in all-to-all connectivity.
\end{abstract}

\vspace{0.5cm}


\vspace{0.5cm}
]

\section{Introduction}

Modern portfolio optimization is fundamentally governed by the imperative to balance expected asset returns against systemic market risk. Traditional quantitative models, pioneered by Harry Markowitz in Modern Portfolio Theory (MPT) \cite{Markowitz1952}, rely heavily on standard, symmetric covariance matrices to measure this risk. In the Markowitz framework, risk is treated simply as the standard deviation (volatility) of asset returns. However, empirical market dynamics over the last two decades punctuated by flash crashes, global liquidity crises, and catastrophic losses have repeatedly exposed the inadequacy of naive symmetric risk metrics. 

Financial markets exhibit "fat tails", meaning extreme, asymmetric downturns occur far more frequently than a standard Gaussian normal distribution would predict. During these tail risk events, previously uncorrelated assets often crash simultaneously, rendering traditional diversification strategies highly ineffective. This reality dictates the integration of advanced tail risk metrics, such as Conditional Value at Risk (CVaR) \cite{Rockafellar2000, Woerner2019}, to protect capital during periods of severe market stress. 

However, incorporating both Mean Variance and CVaR into a single discrete optimization framework generates a highly dense, nonlinear energy landscape. As the universe of investable assets grows, the computational complexity scales exponentially. The process of identifying the optimal combination of discrete asset lots transforms the problem into an NP-Hard combinatorial challenge that severely strains classical algorithmic limits \cite{Orus2019, Herman2022, Pistoia2021}.

Quantum optimization algorithms, notably the Quantum Approximate Optimization Algorithm (QAOA) \cite{Farhi2014} and parameterized Variational Quantum Eigensolvers (VQE) \cite{Cerezo2021}, have emerged as leading candidates to resolve these computationally prohibitive Quadratic Unconstrained Binary Optimization (QUBO) formulations \cite{Lucas2014, Rosenberg2016, Egger2020}. 

Further, a severe disconnect currently exists between theoretical quantum algorithms and available physical hardware. While the analytical performance is guaranteed for algorithms like QAOA at infinite theoretical limits, their physical realization on Noisy Intermediate Scale Quantum (NISQ) architectures \cite{Preskill2018} requires complex and highly lossy circuit compilation. Specifically, embedding dense financial problems onto IBM's sparsely connected planar heavy hex lattices \cite{Chamberland2020} requires extensive nonlocal qubit routing \cite{Weidenfeller2022, Brandhofer2022}. 

This paper investigates the hardware level viability of overcoming these limitations. It seeks to determine whether substituting a mathematically rigid, problem specific ansatz (WS-QAOA) with a highly constrained, hardware isomorphic structure (HE-VQNN) can successfully bypass the decoherence limits induced by physical routing overhead, without sacrificing the algorithm's mathematical ability to accurately find the optimal portfolio.

\section{Theoretical Framework}

\subsection{Value at Risk and Exact CVaR Formulation}

To quantify tail risk mathematically, financial institutions frequently utilize Value at Risk (VaR). VaR represents the maximum expected loss $L(x, r)$ for a specific portfolio weights vector $x$ and a stochastic asset return vector $r$, over a defined time horizon at a given confidence level $\alpha \in (0,1)$ (typically set at 95\% or 99\%). VaR is defined  as the threshold where the probability of a loss exceeding that threshold is exactly $1 - \alpha$:
\begin{equation}
    \text{VaR}_{\alpha}(x) = \min \left\{ \zeta \in \mathbb{R} : \mathbb{P}(L(x, r) \le \zeta) \ge \alpha \right\}
\end{equation}

While ubiquitous in the banking industry due to regulatory requirements, VaR is mathematically flawed for complex optimization. It lacks subadditivity, a foundational mathematical property required to justify the risk reducing benefits of asset diversification (i.e., the risk of a combined portfolio should be less than or equal to the sum of individual risks: $Risk(A+B) \le Risk(A) + Risk(B)$). Furthermore, VaR is completely blind to the magnitude of losses that strictly exceed the $\zeta$ threshold; it only identifies where the "tail" begins, providing no information regarding how severe the extreme losses might be.

Conditional Value at Risk (CVaR), also known as Expected Shortfall, resolves these severe geometric limitations by quantifying the expected, probability weighted loss strictly within that worst $(1-\alpha)$ quantile \cite{Woerner2019}. Following the breakthrough continuous formulation established by Rockafellar and Uryasev \cite{Rockafellar2000}, CVaR transforms a complex, nonconvex sorting problem into a convex optimization problem through the introduction of an auxiliary continuous variable $\zeta$:

{\small
\begin{equation}
\begin{aligned}
    F_{\alpha}(x, \zeta) &= \zeta + \frac{1}{1-\alpha} \int \max(0, L(x, y) - \zeta) p(y) dy
\end{aligned}
\end{equation}
}
Where $p(y)$ represents the probability density function of market scenarios. Because CVaR satisfies the axioms of both convexity and coherence, it provides a mathematically stable and globally solvable landscape for large scale risk optimization. 

\subsection{The Quantum Bottleneck: Discretization}

While the Rockafellar-Uryasev formulation is highly efficient for classical linear programming, integrating it directly into a quantum objective function poses a massive architectural challenge \cite{Barkoutsos2020}. To evaluate the continuous integral on digital hardware, it must be discretized over a finite set of $S$ historical market scenarios:
\begin{equation}
    \text{CVaR}_{\alpha}(x) = \min_{\zeta} \left( \zeta + \frac{1}{S(1-\alpha)} \sum_{s=1}^S z_s \right)
\end{equation}
subject to the constraints $z_s \ge L_s(x) - \zeta$ and $z_s \ge 0$, where $z_s$ represents the continuous slack variables tracking the loss in each scenario.

In a discrete binary quantum environment, representing these continuous auxiliary variables necessitates complex fractional binary discretization \cite{Hodson2019, Abbas2023}. The continuous threshold $\zeta$ and every single slack variable $z_s$ must be explicitly mapped to dedicated registers of quantum bits using a binary expansion (e.g., $z_s \approx \sum_{m=-K}^K 2^m y_m$). 

Consequently, for a financial dataset evaluating $S$ historical scenarios with a precision of $P$ bits, the total logical qubit requirement explodes to:
\begin{equation}
    N_{qubits} = N_{assets} + P_{threshold} + (S \times P_{slack})
\end{equation}

To put this into perspective: evaluating a modest 10 asset portfolio over 250 trading days (one standard financial year) at a basic 8 bit precision would demand $10 + 8 + (250 \times 8) = 2,018$ perfectly coherent, error corrected logical qubits. This auxiliary variable bottleneck renders the exact continuous CVaR formulation completely and physically uncomputable on current NISQ hardware, which is strictly limited to around 100 to 150 inherently noisy physical qubits.

\subsection{The Novel Approach: Hybrid Proxy Matrix}

To successfully circumvent the auxiliary qubit bottleneck and preserve a strictly efficient $1:1$ asset to qubit mapping (where $N_{qubits} = N_{assets}$), this study introduces and implements a classical quantum hybrid preprocessing heuristic \cite{Mugel2022}. Rather than forcing the fragile quantum processor to dynamically evaluate the threshold $\zeta$ and track hundreds of slack variables using thousands of gates, the mathematical complexity is forcefully integrated out during a classical preprocessing phase.

The exact discrete CVaR objective relies on minimizing the linear sum of slack variables $z_s$, governed by the mathematical constraint $z_s = \max(0, L_s - \zeta)$. A strictly positive slack variable ($z_s > 0$) mathematically dictates that the portfolio loss for a specific scenario $s$ has actively breached the VaR threshold. To completely eliminate the need for these auxiliary quantum registers, our hybrid framework abandons the dynamic quantum calculation of $z_s$. Instead, we isolate the specific historical scenarios that would theoretically produce $z_s > 0$ utilizing a classical CPU prior to QPU execution.

Given a historical return matrix $\mathbf{R} \in \mathbb{R}^{T \times N}$ (where $T$ represents the total trading days), the proxy portfolio loss for each historical day $t$ under an equal weight assumption is computed as:
\begin{equation}
    L_t = -\sum_{i=1}^N R_{t,i}
\end{equation}

Sorting the resultant set $\{L_t\}$ in ascending order allows for the exact empirical derivation of the static VaR decision boundary at the specified $(1-\alpha)$ confidence level:
\begin{equation}
    \zeta^* = \text{Quantile}\left(\{L_t\}_{t=1}^T,\; \alpha\right)
\end{equation}

Once this classical threshold is established, normal market days are mathematically discarded. The algorithm exclusively isolates the specific set of days ($\mathcal{T}_{tail}$) where the recorded portfolio loss strictly exceeded the threshold (mirroring the precise condition where the slack variable $z_s > 0$):
\begin{equation}
    \mathcal{T}_{tail} = \left\{ t \in \{1,\ldots,T\} : L_t > \zeta^* \right\}
\end{equation}

Instead of enforcing a linear penalty ($\sum z_s$) across these scenarios which inherently triggers the qubit explosion, we extract the corresponding return submatrix $\mathbf{R}_{tail} \in \mathbb{R}^{|\mathcal{T}_{tail}| \times N}$. To computationally penalize these simultaneous asset crashes natively within a QUBO structure, we compute the sample covariance matrix strictly over this crash regime:
\begin{equation}
    {\Sigma}_{tail} = \frac{1}{|\mathcal{T}_{tail}| - 1} \left(\mathbf{R}_{tail} - \bar{\mathbf{R}}_{tail}\right)^T \left(\mathbf{R}_{tail} - \bar{\mathbf{R}}_{tail}\right)
\end{equation}

This $\Sigma_{tail}$ matrix acts as a quadratic, QPU native surrogate for the exact linear CVaR objective. It intrinsically captures the severe, nonlinear correlation spikes that materialize specifically during systemic panic exactly the dynamics that standard Markowitz covariance ignores. 

The standard, global Markowitz risk matrix ($\Sigma$) is then hybridized with this localized CVaR proxy matrix via scalar addition:
\begin{equation}
    {\Sigma}_{hybrid} = \frac{q_1}{2}{\Sigma} + \frac{q_2}{2}{\Sigma}_{tail}
\end{equation}
where the scalars $q_1$ and $q_2$ act as tunable penalty weights for general market volatility and severe downside tail risk, respectively. 

By pushing the continuous mathematical complexity into the classical preprocessing phase, the quantum processor is presented with a standard, highly compressed quadratic objective. For a binary portfolio vector $x \in \{0,1\}^N$, expected return vector $\mu$, and a cardinality constraint enforcing a strict asset budget $B$, the constrained optimization is mapped to an unconstrained objective function via a quadratic penalty factor $\eta$:

{\small
\begin{equation}
\begin{aligned}
    \min_{x} \Bigg[ & -\lambda \sum_{i=1}^N \mu_i x_i + \sum_{i,j=1}^N x_i \Sigma_{ij, hybrid} x_j \\
    & + \eta \left( \sum_{i=1}^N x_i - B \right)^2 \Bigg]
\end{aligned}
\end{equation}
}

This formulation represents a significant methodological novelty. It successfully injects deep CVaR tail risk awareness into the quantum search space without requiring a single auxiliary qubit, allowing advanced financial models to be evaluated efficiently on near term sparse quantum devices.

\subsection{Transformation to Ising Hamiltonian}

To execute the objective on a quantum processing unit (QPU), the QUBO formulation must be algebraically translated into a physical Ising Hamiltonian. By expanding the squared budget penalty constraint and leveraging the binary identity $x_i^2 = x_i$ (since $x \in \{0,1\}$), the terms are systematically isolated into linear and quadratic coefficients:

{\small
\begin{equation}
\begin{aligned}
    \left( \sum_{i=1}^N x_i - B \right)^2 &= \sum_{i=1}^N x_i^2 + 2\sum_{i<j} x_i x_j \\
    &\quad - 2B\sum_{i=1}^N x_i + B^2
\end{aligned}
\end{equation}
}

The resulting elements of the target QUBO matrix $Q$ are thereby defined as:
\begin{equation}
    Q_{ii} = -\lambda \mu_i + {\Sigma}_{ii, hybrid} + \eta(1 - 2B)
\end{equation}
\begin{equation}
    Q_{ij} = {\Sigma}_{ij, hybrid} + 2\eta \quad (\text{for } i \neq j)
\end{equation}

The QUBO matrix is ultimately mapped to an Ising Hamiltonian $H_C$ by transforming the classical binary variables $x_i \in \{0,1\}$ to quantum Pauli-Z spin operators $Z_i \in \{+1, -1\}$ via the standard physics algebraic mapping $x_i = \frac{1}{2}(I - Z_i)$ \cite{Lucas2014}. Substituting this mapping yields the final, executable problem Hamiltonian:
\begin{equation}
    H_C = \sum_{i=1}^N h_i Z_i + \sum_{i<j}^N J_{ij} Z_i Z_j
\end{equation}
where $h_i$ represents the local magnetic field bias applied to physical qubit $i$, and $J_{ij}$ represents the critical $Z \otimes Z$ coupling interaction strength between qubits $i$ and $j$. 

\subsection{The Density Problem: Novelty of the Benchmark Scale}

Standard Markowitz models utilizing carefully selected, diverse, uncorrelated assets can occasionally yield mathematically sparse $J_{ij}$ matrices, which are relatively easy for quantum computers to map and solve. However, the ${\Sigma}_{hybrid}$ matrix required to accurately reflect CVaR tail risk is fundamentally and unavoidably dense. During extreme market events, systemic risk causes previously uncorrelated assets to move in tandem, resulting in nonzero covariance values across nearly the entire interaction matrix.

\begin{figure}[ht]
    \centering
    \fbox{\includegraphics[width=\linewidth]{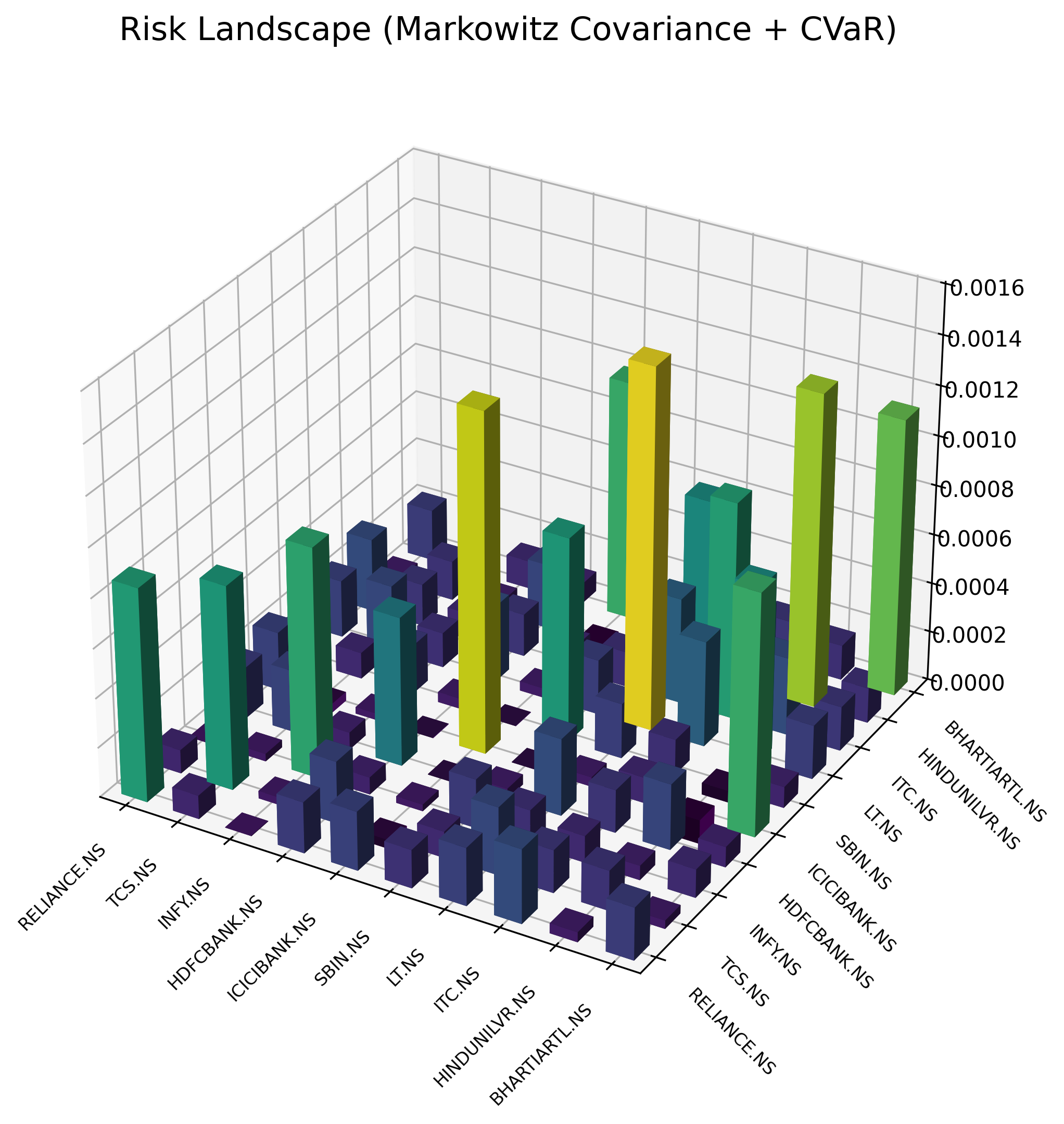}}
    \caption{3D visualization of the hybrid Markowitz CVaR risk landscape (${\Sigma}_{hybrid}$) for the evaluated 10 asset NIFTY 50 portfolio. The dense distribution of nonzero off diagonal elements mandates all to all logical connectivity.}
    \label{fig:risk_landscape}
\end{figure}

As visually verified in Figure \ref{fig:risk_landscape}, evaluating real world market data across major indices generates a heavily populated covariance matrix. In graph theory, this extreme density forces the corresponding QUBO representation to form a complete, nonplanar graph ${K_N}$ \cite{Harrigan2021}, meaning every individual asset must dynamically interact with every other asset during the quantum calculation. 

A 10 asset scale was deliberately selected as the benchmark baseline for this study because it represents a critical hardware threshold. Mathematically, a complete graph $K_{N}$ requires exactly $N(N-1)/2$ unique pairwise connections. Therefore, 10 assets require 45 connections, and 16 assets require 120. On IBM's physical heavy hex topology, the maximum theoretical connectivity degree of any single physical qubit is strictly limited to three (or two at the edges) \cite{Chamberland2020}.
 
Embedding 45 dense, all to all interactions onto this highly sparse physical architecture necessitates massive nonlocal routing via intermediary SWAP networks \cite{Bravyi2022}. Therefore, 10 assets represent the precise mathematical inflection point where physical hardware routing overhead begins to severely overwhelm the processor's limited coherence times. Testing at this exact scale serves as an optimal stress test for evaluating the true algorithmic resilience of different quantum approaches in the NISQ era.

\section{Quantum Algorithmic Approaches}

\subsection{Hardware Efficient VQNN (HE-VQNN)}

To artificially decouple the required quantum circuit depth from the high mathematical density of the financial problem, the HE-VQNN algorithm utilizes an alternating layered ansatz \cite{Kandala2017} constructed exclusively from native hardware operations. Instead of attempting to map the complex $K_N$ graph directly to the hardware, the trial state $\ket{\psi(\vec{\theta})}$ is prepared through sequential layers of parameterized single qubit rotations $U_{rot}$ and fixed two qubit entanglement operations $U_{ent}$:
\begin{equation}
    \ket{\psi(\vec{\theta})} = \prod_{l=1}^{L} \left( U_{ent} U_{rot}(\vec{\theta}_l) \right) \ket{0}^{\otimes N}
\end{equation}
where $U_{rot}(\vec{\theta}_l) = \bigotimes_{i=1}^N R_y(\theta_{i,l})$. The system begins in the ground state $\ket{0}^{\otimes N}$, and the $R_y$ gates apply parameterized rotations around the Y axis of the Bloch sphere, creating a tunable quantum superposition.

\begin{figure}[ht]
    \centering
    \makebox[2pt][c]{%
    \fbox{\includegraphics[width=1.00\linewidth]{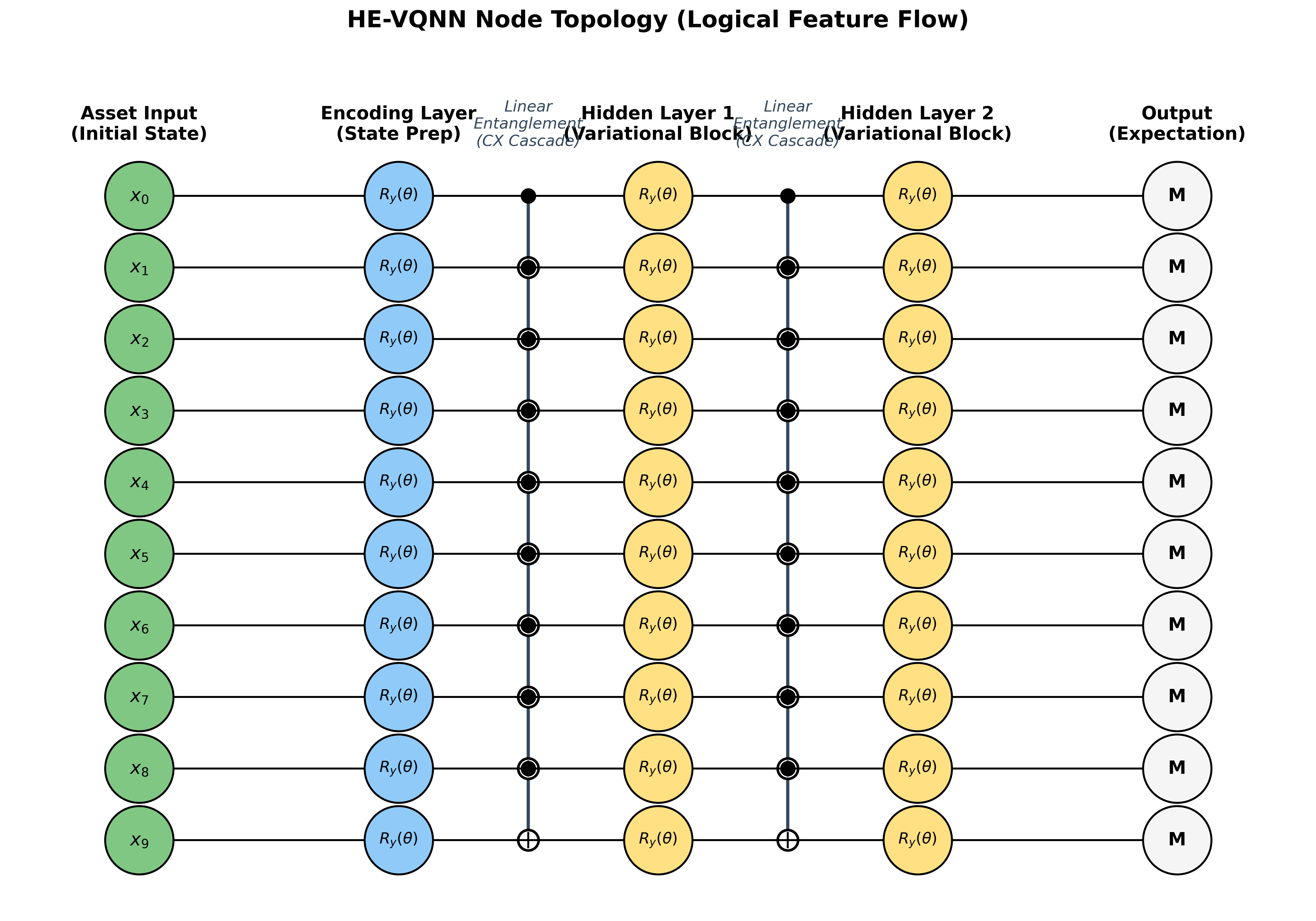}}%
    }
    \caption{HE-VQNN Node Topology: The logical feature flow dictates linear entanglement cascades, circumventing the need for all to all connectivity.}
    \label{fig:logical_topology}
\end{figure}

Crucially, as shown in Figure \ref{fig:logical_topology}, the entanglement layer $U_{ent}$ utilizes a strictly linear cascade of controlled NOT (CX) gates that only connect adjacent index qubits ($q_i \to q_{i+1}$). 

Because the mathematical structure of $U_{ent}$ intentionally limits entanglement to a 1D chain, the resulting circuit depth remains highly static and scales linearly $\mathcal{O}(L \cdot N)$. This design choice purposefully sacrifices mathematical complexity to ensure that the physical execution time on the quantum processor remains entirely agnostic to the density of the $K_N$ CVaR covariance matrix.

\begin{figure}[ht]
    \centering
    \makebox[0pt][c]{%
        \hspace{0.01\linewidth}\fbox{\includegraphics[width=\linewidth, height = 1.0\linewidth]{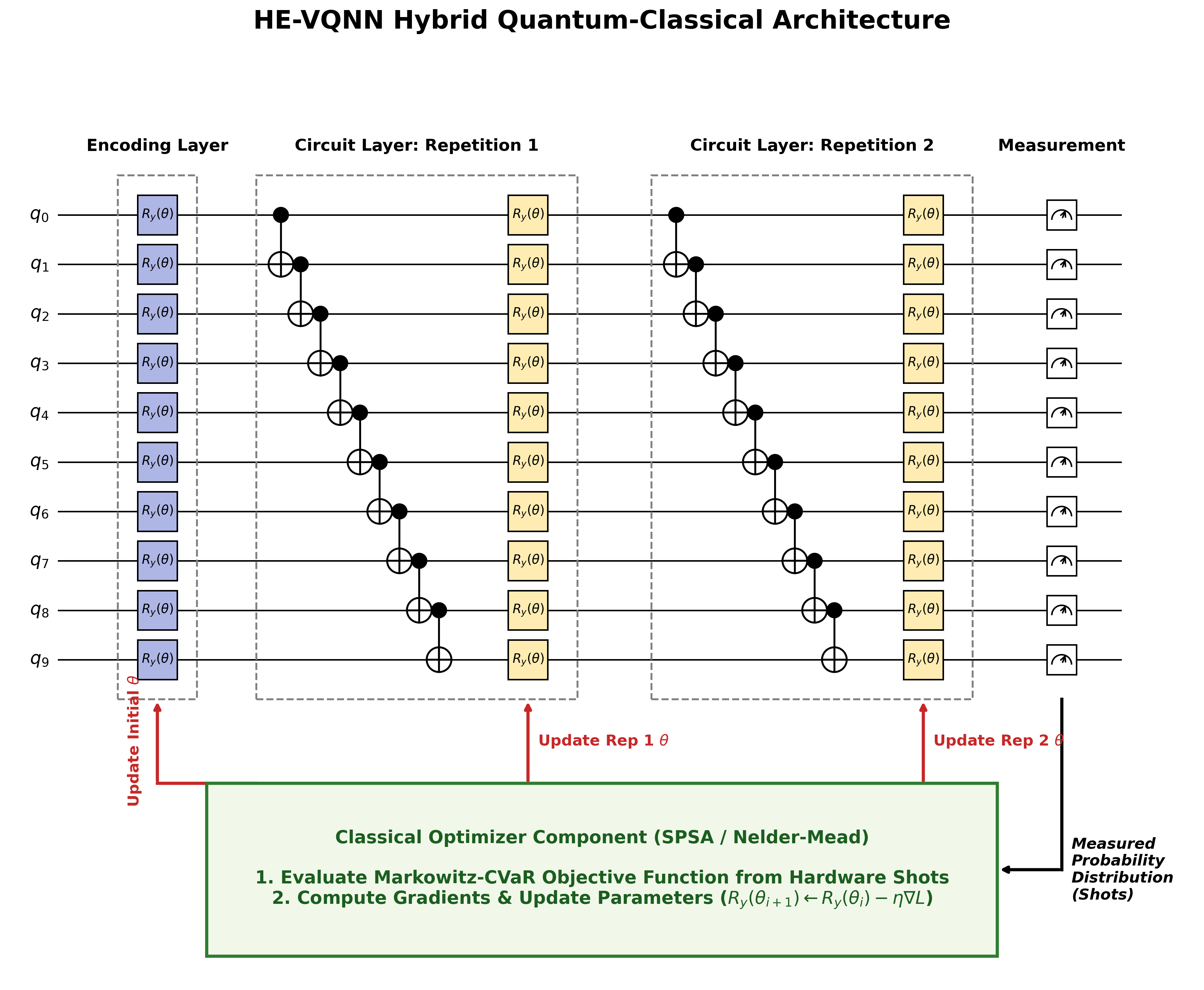}}%
    }
    \caption{Hybrid Quantum Classical architecture for HE-VQNN, detailing the feedback control loop between physical measurement shots and the classical parameter optimization routine \cite{Biamonte2017}.}
    \label{fig:hybrid_arch}
\end{figure}

\subsection{Warm Start QAOA (WS-QAOA)}

Unlike the hardware agnostic HE-VQNN, Standard QAOA attempts to encode the specific problem geometry directly into the quantum state via phase separator unitaries $e^{-i {\gamma} H_C}$ \cite{Farhi2014}. However, standard QAOA begins in an equal superposition state, which requires exploring the vast $2^N$ combinatorial search space entirely from scratch. 

To restrict this search space and theoretically accelerate convergence, Warm Start QAOA \cite{Egger2021} initializes the quantum system in a biased state derived from a classical continuous relaxation of the QUBO problem (yielding a classical probability vector $x^* \in [0,1]^N$). 

Instead of an equal superposition, the initial quantum state is shifted to favor the classical approximation:
\begin{equation}
    \ket{\phi^*} = \bigotimes_{i=1}^N \left( \sqrt{1-x_i^*}\ket{0} + \sqrt{x_i^*}\ket{1} \right)
\end{equation}

Furthermore, the standard Pauli X mixer used in base QAOA is replaced with a custom mixer Hamiltonian $H_M$, which is projected specifically into the subspace defined by the classical relaxation $x^*$:
\begin{equation}
    H_M = \sum_{i=1}^N \left( \cos(\theta_i) X_i + \sin(\theta_i) Y_i \right)
\end{equation}
where $\theta_i = 2 \arcsin(\sqrt{x_i^*})$. This ensures that the quantum algorithm is only exploring localized regions of the Hilbert space that are likely to contain the optimal portfolio.

To clarify the physical implementation shown in the circuit diagram, we must consider the unitary evolution generated by this custom mixer, $U_M{(\beta)} = e^{-i {\beta} H_M}$. For an individual qubit $i$, the mixer term is geometrically equivalent to a Pauli X operator rotated around the Y axis by the angle $\theta_i$. Mathematically, this basis transformation is expressed as:
\begin{equation}
    \cos(\theta_i) X_i + \sin(\theta_i) Y_i = R_Y(\theta_i) X_i R_Y(-\theta_i)
\end{equation}

Substituting this into the unitary evolution yields:
\begin{equation}
    U_{M,i}{(\beta)} = \exp \left( -i \beta \left[ R_Y(\theta_i) X_i R_Y(-\theta_i) \right] \right)
\end{equation}

Exploiting the standard matrix exponential property $e^{UAU^\dagger} = U e^A U^\dagger$, the $R_Y$ rotation matrices can be factored outside the exponential:
\begin{equation}
    U_{M,i}{(\beta)} = R_Y(\theta_i) e^{-i \beta X_i} R_Y(-\theta_i)
\end{equation}

Because the standard single qubit X rotation gate is defined as $R_X{(\alpha)} = e^{-i \frac{\alpha}{2} X}$, the core exponential term $e^{-i {\beta} X_i}$ corresponds exactly to an $R_X(2{\beta})$ gate. Therefore, the continuous time evolution of the projected mixer translates to the following discrete, hardware native gate sequence:
\begin{equation}
    U_{M,i}{(\beta)} = R_Y(\theta_i) R_X(2\beta) R_Y(-\theta_i)
\end{equation}

This derivation perfectly aligns with the circuit architecture, demonstrating how the quantum state is temporarily rotated out of the warm start basis by $R_Y(-{\theta}_i)$, mixed using a standard X rotation, and subsequently rotated back into the restricted subspace by $R_Y({\theta}_i)$.

\begin{figure}[htbp!]
    \centering
    \fbox{\includegraphics[width=1.0\linewidth, height = 0.65\linewidth, scale  = 1.6]{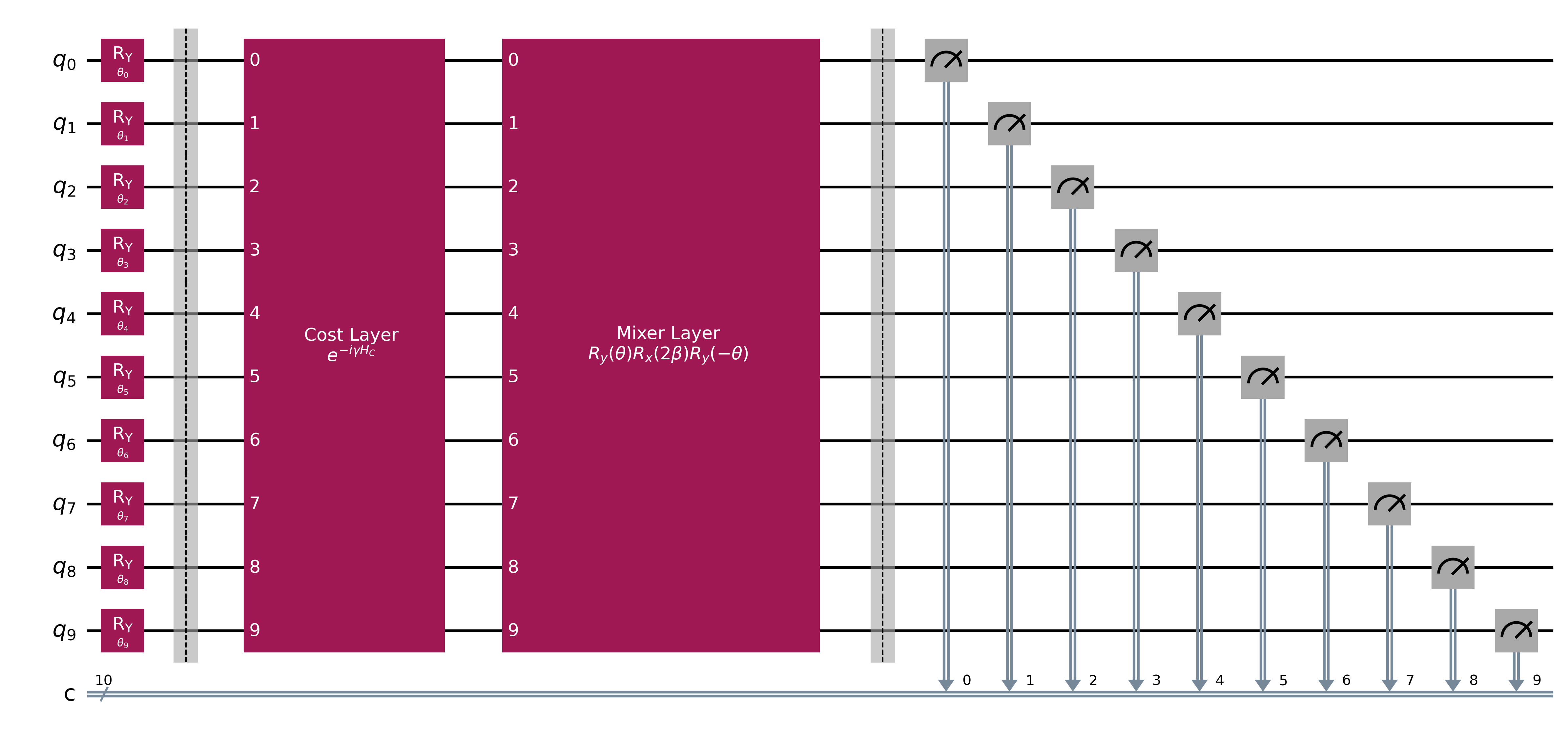}} 
    \caption{Simulator level abstract quantum circuit diagram for WS-QAOA, highlighting the custom initial state preparations and complex mixer blocks.}
    \label{fig:ws_sim_circuit}
\end{figure}

While WS-QAOA provides robust, mathematically rigorous convergence guarantees at higher depths, its physical implementation presents a massive challenge. The exact physical execution of the phase separator operator $e^{-i {\gamma} H_C}$ dictates that the quantum hardware must physically implement a $Z \otimes Z$ interaction gate (typically built using two physical CNOT gates and an $R_z$ rotation) for every single nonzero $J_{ij}$ term in the Hamiltonian. As dictated by the dense connectivity requirement shown earlier in Figure \ref{fig:risk_landscape}, attempting to force this mathematical requirement onto a sparse hardware processor guarantees the generation of a massive, deep grid of SWAP operations upon transpilation to the hardware.

\section{Classical Optimization Strategies}

\begin{figure}[ht]
    \centering
    \makebox[0pt][c]{%
        \hspace{0.02\linewidth}\fbox{\includegraphics[width=1.02\linewidth, height = 0.8\linewidth, scale = 1.6]{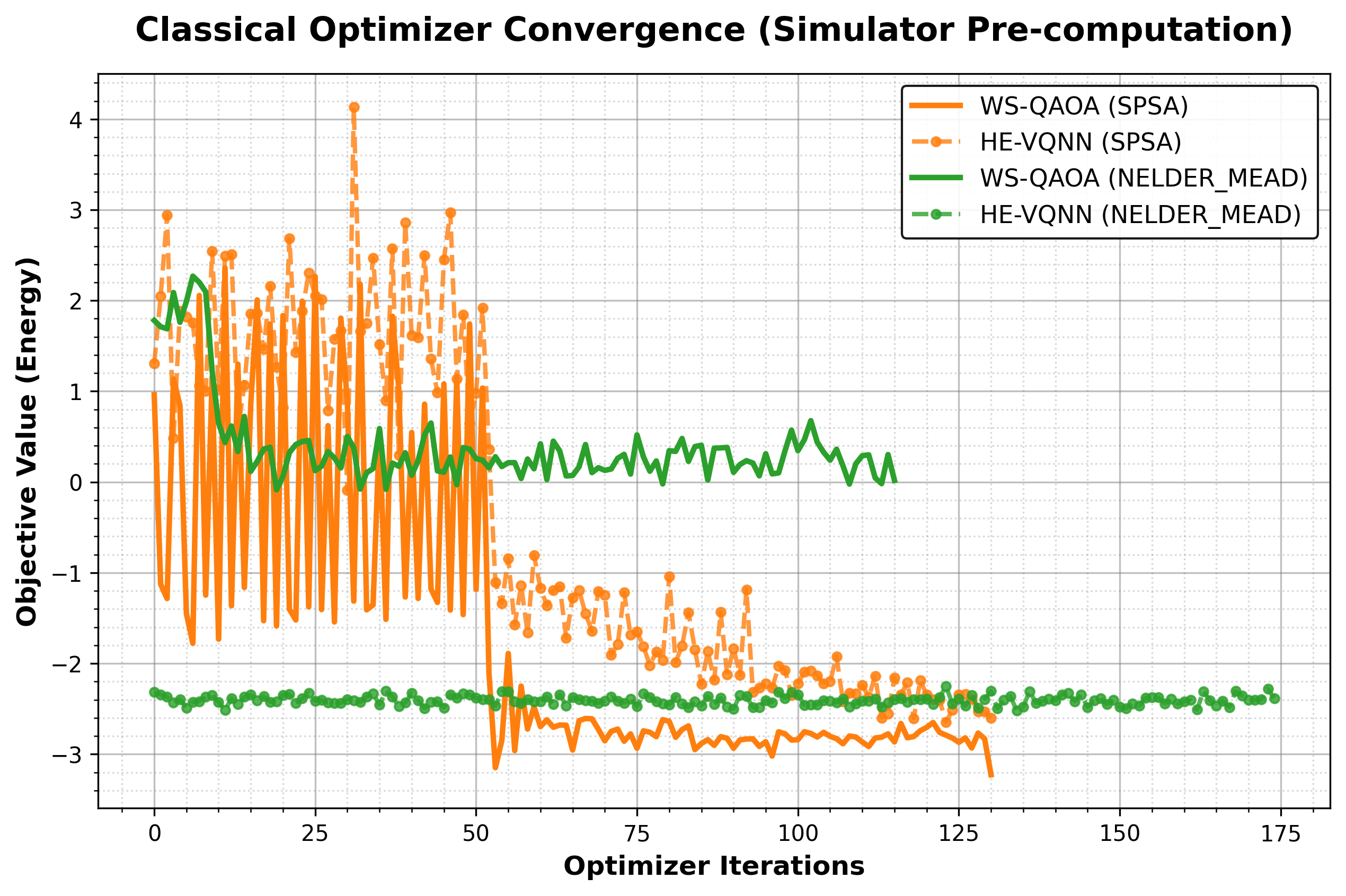}}%
    }
    \caption{Optimizer convergence history for WS-QAOA and HE-VQNN simulated prior to hardware deployment. Evaluated utilizing SPSA and Nelder Mead routines.}
    \label{fig:convergence}
\end{figure}

Variational quantum algorithms require external classical optimizers to iteratively adjust the parameters $\vec{\theta}$ of the quantum ansatz based on measurement feedback from the QPU. In this study, the performance of the quantum circuits is benchmarked utilizing two distinct classical optimization paradigms. Figure \ref{fig:convergence} tracks the optimizer performance during the initial noiseless precomputation phase on a statevector simulator.

\subsection{Simultaneous Perturbation Stochastic Approximation (SPSA)}

SPSA \cite{Spall1992} is a stochastic gradient descent approximation method specifically engineered for noisy, error prone objective functions. Traditional finite difference methods require $2N$ distinct circuit evaluations to approximate the gradient for a model with $N$ parameters. For a 10 asset portfolio, this means 20 evaluations per optimization step, which is overwhelmingly slow and resource intensive on real quantum hardware. 

Conversely, SPSA estimates the overall gradient vector $\hat{g}_k(\vec{\theta}_k)$ using only two quantum circuit evaluations per iteration, regardless of parameter size. The gradient is approximated via a randomly generated perturbation vector $\Delta_k$:

{\small
\begin{equation}
\begin{aligned}
    \hat{g}_k(\vec{\theta}_k) &= \frac{\mathcal{L}(\vec{\theta}_k + c_k \Delta_k) - \mathcal{L}(\vec{\theta}_k - c_k \Delta_k)}{2 c_k} \\
    &\quad \times \Delta_k^{-1}
\end{aligned}
\end{equation}
}

where $c_k$ dictates the step size. As demonstrated by the highly erratic but generally descending blue line in Figure \ref{fig:convergence}, SPSA exhibits high variance. However, it maintains strong mathematical resilience against the inherent thermal relaxation and gate noise present in physical superconducting hardware, allowing the algorithm to actively bounce out of local minima traps effectively.

\subsection{Nelder-Mead Simplex Method}

The Nelder Mead method is a gradient free heuristic that utilizes a geometric simplex of $N+1$ points in an $N$ dimensional space to navigate the energy landscape. The algorithm iteratively updates the simplex via geometric 
, expansion, contraction, and shrink operations to rapidly locate the minimum objective value. While this method is highly effective in stable, noiseless simulated environments (serving as the smooth, rapid exact baseline shown in Figure \ref{fig:convergence}), it lacks noise awareness. Because it relies on exact point evaluations, it frequently fails to converge when subjected to the unpredictable stochastic shot noise and device drifts of physical NISQ hardware. 

\section{Methodology}

Financial data was sourced from the NIFTY 50 index across a multiyear horizon specifically tailored to capture periods of macroeconomic instability and high volatility. This ensures the dataset provides a valid and challenging tail risk distribution for the CVaR objective to evaluate. The transpilation and execution pipeline targeted the \texttt{ibm\_fez} 127 qubit heavy hex processor using the Qiskit framework \cite{Qiskit2024}. To minimize physical depth during hardware execution, a stochastic Sabre routing algorithm was integrated with aggressive peephole optimization. Barrier removal and unitary synthesis at a strict $0.98$ approximation degree were implemented to actively compress redundant local unitary matrices before sending the microwave pulse schedules to the QPU.

\section{Results and Discussion}

\subsection{Transpilation Metrics and Hardware Constraints}

Prior to hardware execution, both algorithmic ansatzes were fully transpiled from their abstract logical representations to the \texttt{ibm\_fez} native Instruction Set Architecture (ISA) for the 10 asset baseline scale. Table \ref{tab:hardware_metrics} details the precise physical resource requirements extracted post transpilation. As theoretically predicted by the high density of the CVaR matrix, the WS-QAOA circuit depth exploded significantly compared to the hardware efficient model. 

\begin{table*}[t!]
\centering
\caption{Transpiled Hardware Execution Metrics for \texttt{ibm\_fez} (10 Assets)}
\label{tab:hardware_metrics}
\renewcommand{\arraystretch}{1.4}
\begin{tabular}{@{}lcc@{}}
\toprule
\textbf{Hardware Metric} & \textbf{HE-VQNN (2 Hidden Layers)} & \textbf{WS-QAOA (Custom Mixer, reps=1)} \\ 
\midrule
Total Abstract Qubits & 10 & 10 \\
Physical Qubits Mapped & 10 & 10 \\
Abstract Circuit Depth & 15 & 22 \\
ISA Transpiled Depth (Hardware) & 51 & 301 \\
Total Physical Instructions (Gate Count) & 187 & 739 \\
Single Qubit Gates ($R_z, \sqrt{X}, X$) & 149 & 565 \\
Nonlocal Routing Gates (CNOT/ECR) & 18 & 164 \\
Estimated Circuit Duration ($\mu$s) & 5.10 $\mu$s & 30.10 $\mu$s \\
Hardware Median $T_1$ Decay Time ($\mu$s) & 147.43 $\mu$s & 147.43 $\mu$s \\
Hardware Median $T_2$ Dephasing Time ($\mu$s) & 109.80 $\mu$s & 109.80 $\mu$s \\
\textbf{Estimated Success Prob (ESP - $T_2$ limit)} & \textbf{95.46\%} & \textbf{76.02\%} \\ 
\bottomrule
\end{tabular}
\end{table*}

\begin{figure}[htbp!]
    \centering
    \fbox{\includegraphics[width=1.03\linewidth, height = 0.68\linewidth, scale  = 1.8]{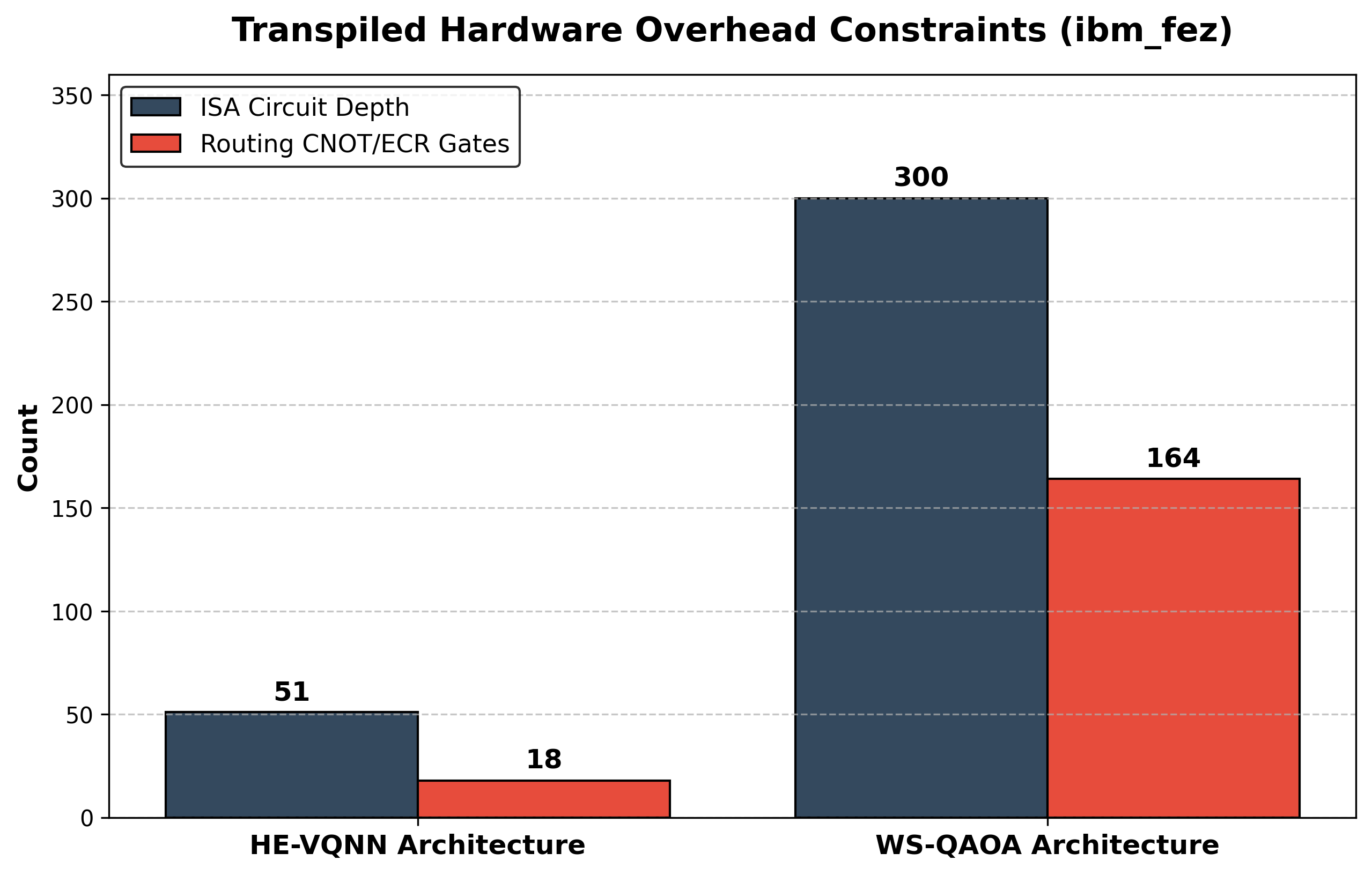}}
    \caption{Comparison of transpiled hardware overhead constraints on \texttt{ibm\_fez}. WS-QAOA exhibits an explosive increase in both circuit depth and nonlocal routing operations compared to the hardware efficient ansatz.}
    \label{fig:overhead_bar}
\end{figure}

Figure \ref{fig:overhead_bar} visually emphasizes this stark contrast in physical resource scaling. While the HE-VQNN successfully maintains a shallow hardware depth of 51 utilizing only 18 routing gates, the WS-QAOA requires a staggering ISA depth of 301 and 164 nonlocal routing gates merely to navigate the heavy hex lattice for a single repetition layer. 

The estimated physical execution duration for WS-QAOA (30.10 $\mu$s) consumes a massive fraction of the processor's median $T_2$ dephasing time (109.80 $\mu$s). Because quantum states naturally decay over time due to environmental interaction, this mathematical ratio establishes a strong theoretical expectation for a severe loss of hardware signal fidelity prior to measurement. Conversely, the strictly linear, nearest neighbor design of the HE-VQNN circuit naturally aligns with the hardware lattice, allowing it to execute rapidly (5.10 $\mu$s) and preserving a significantly higher theoretical Estimated Success Probability (ESP) threshold of 95.46\%.

\subsection{Hardware Execution and The Expressibility Trap}

The bound circuits were executed directly on the \texttt{ibm\_fez} QPU utilizing 4,096 measurement shots. Using the precomputed optimal SPSA parameters, the evaluated physical hardware expected energy ($E_{hw}$) was benchmarked against the exact global minimum energy ($E_{exact}$) computed via an exact classical NumPy eigensolver to evaluate true optimization accuracy.

\begin{figure}[htbp]
    \centering
    \fbox{\includegraphics[width=1.0\linewidth, height = 0.65\linewidth, scale  = 1.8]{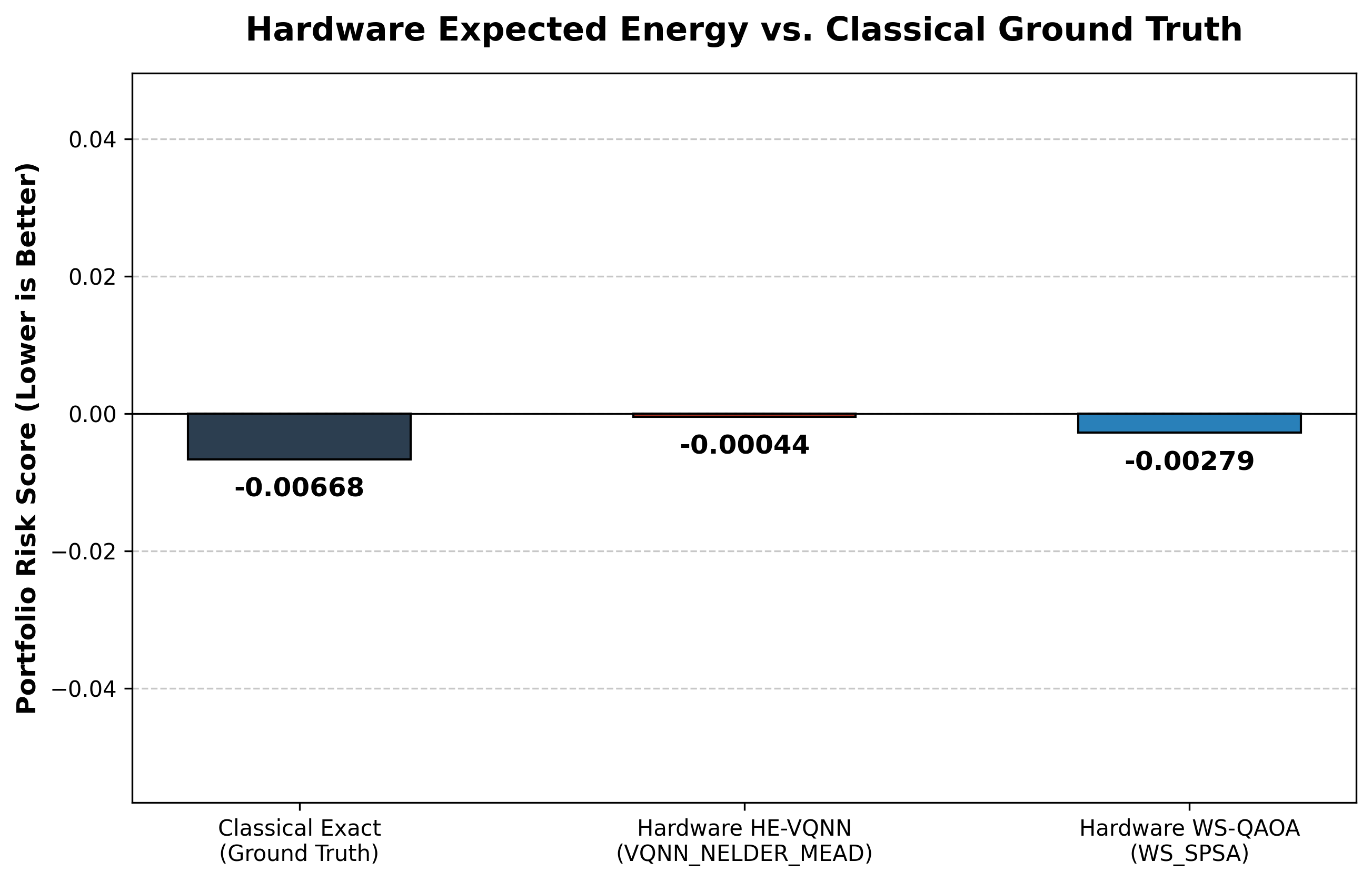}}
    \caption{Absolute energy comparison between evaluated physical hardware states and the theoretical classical ground truth.}
    \label{fig:energy_bar}
\end{figure}

\begin{figure*}[t!]
    \centering
    \begin{subfigure}[b]{0.48\textwidth}
        \centering
        \fbox{\includegraphics[width=\textwidth, height = 0.65 \linewidth]{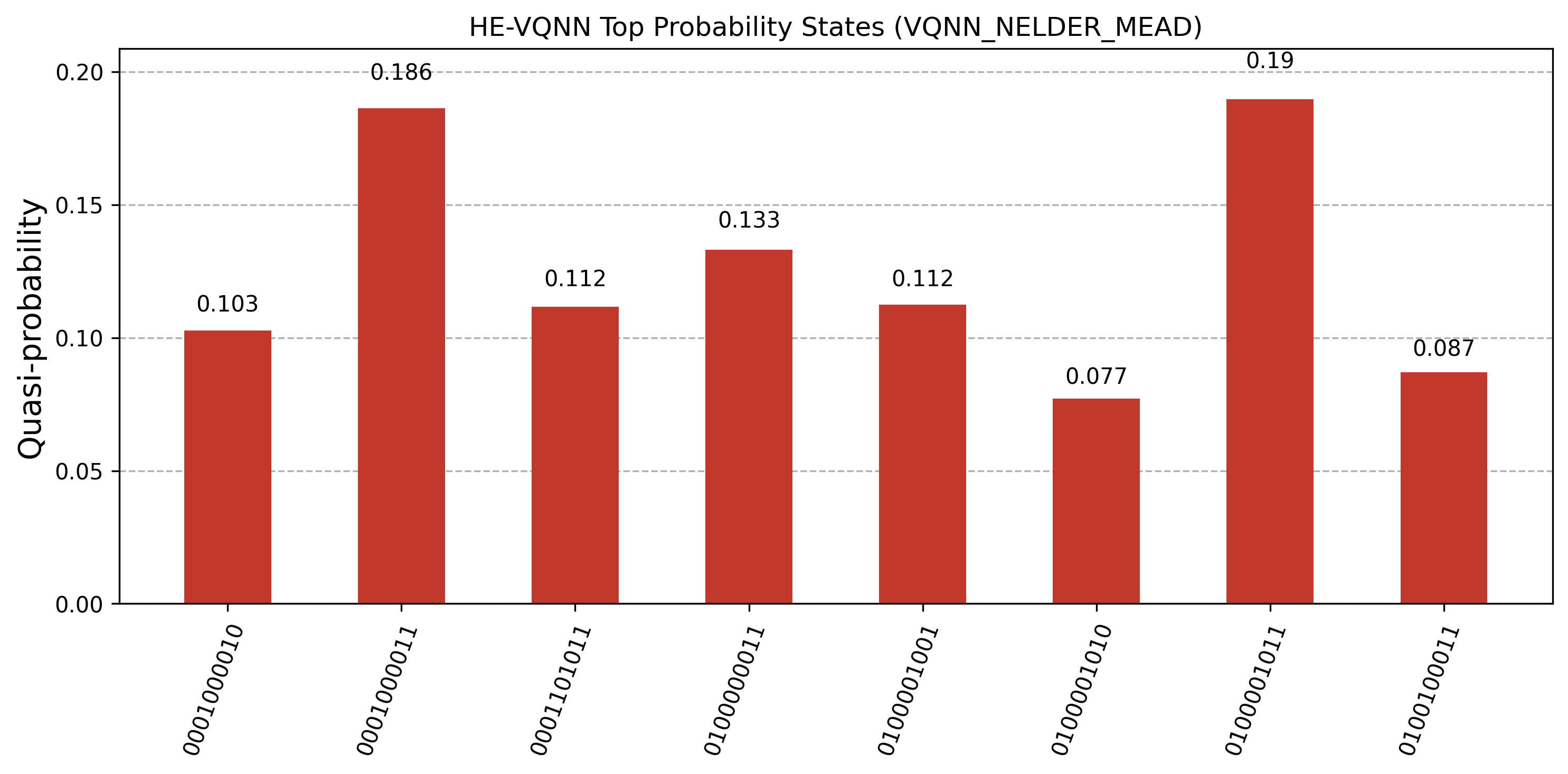}}
        \caption{HE-VQNN state distribution.}
        \label{fig:vqnn_hist}
    \end{subfigure}
    \hfill
    \begin{subfigure}[b]{0.48\textwidth}
        \centering
        \fbox{\includegraphics[width=\textwidth, height = 0.65\linewidth]{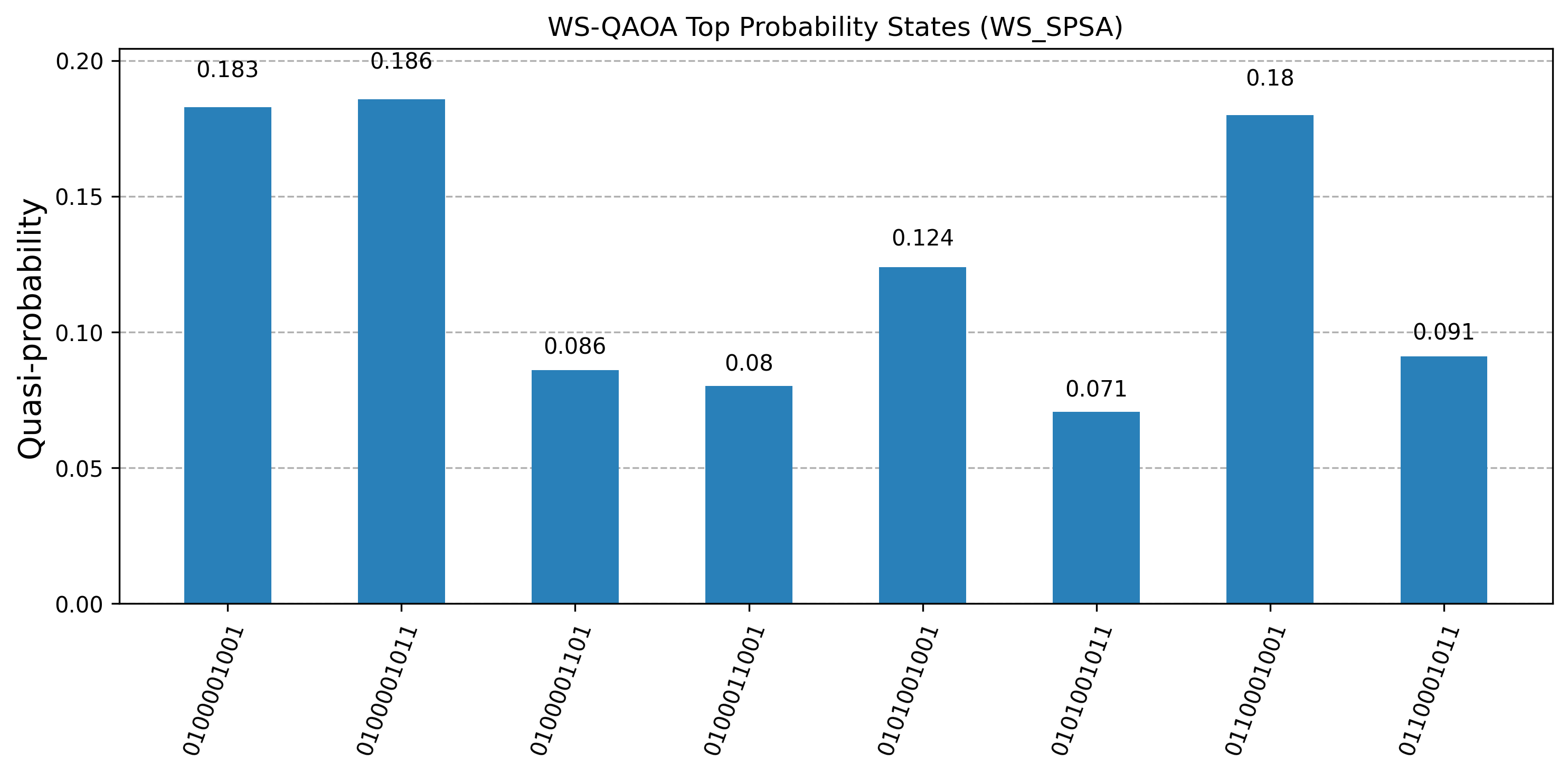}}
        \caption{WS-QAOA state distribution.}
        \label{fig:ws_hist}
    \end{subfigure}
    \caption{Measured state distributions on \texttt{ibm\_fez}. (a) HE-VQNN preserves hardware fidelity but fails to isolate the optimal state due to poor mathematical ansatz expressibility. (b) WS-QAOA maps the geometric problem accurately, but its massive required physical routing degrades the amplitude distribution entirely into static noise.}
    \label{fig:histograms}
\end{figure*}

The quasi probability state distributions extracted from the physical processor reveal two entirely distinct, algorithm specific failure modes. As illustrated in Figure \ref{fig:histograms}(a), despite completing execution well within the hardware coherence limits (indicated by its 95.46\% ESP meaning the quantum state survived the physical gate execution without succumbing to complete thermal decoherence), the HE-VQNN struggled significantly to isolate the true optimal asset combination. Instead of a sharp, distinct probability peak at the correct portfolio configuration, it yielded a relatively flat, high variance probability distribution. 

This specific failure mode highlights a severe theoretical algorithmic limitation. While a strictly linear, nearest neighbor entanglement topology successfully avoids hardware routing noise, it fundamentally lacks the mathematical \emph{expressibility} \cite{Sim2019, Holmes2022, Cincio2021} required to capture the highly correlated, dense multi asset interactions dictated by the CVaR tail risk objective. Because financial assets become highly correlated during systemic market crashes, a simple nearest neighbor ansatz cannot mathematically represent the complex global entanglement required to approximate the true ground state. The neural network easily survives the physical hardware execution, but it is analytically too rigid to model the financial problem's complexity, leading directly to barren plateaus in the optimization landscape \cite{McClean2018}.

Conversely, Figure \ref{fig:histograms}(b) demonstrates that WS-QAOA fails for entirely physical, hardware centric reasons. While mathematically capable of perfectly representing the complex solution space of the dense ${K_{10}}$ graph, the deep SWAP networks required to physically embed that dense matrix onto the sparse, 2D planar hardware layout caused the qubit probability amplitudes to degrade completely into static white noise before measurement could occur. The classical optimizer receives garbage data, rendering the iterative algorithm entirely useless on NISQ devices.

\subsection{Scaling Analysis and The SWAP Penalty}

\begin{figure*}[t!]
    \centering
    \begin{subfigure}[h]{0.48\textwidth}
        \centering
        \fbox{\includegraphics[width=\textwidth]{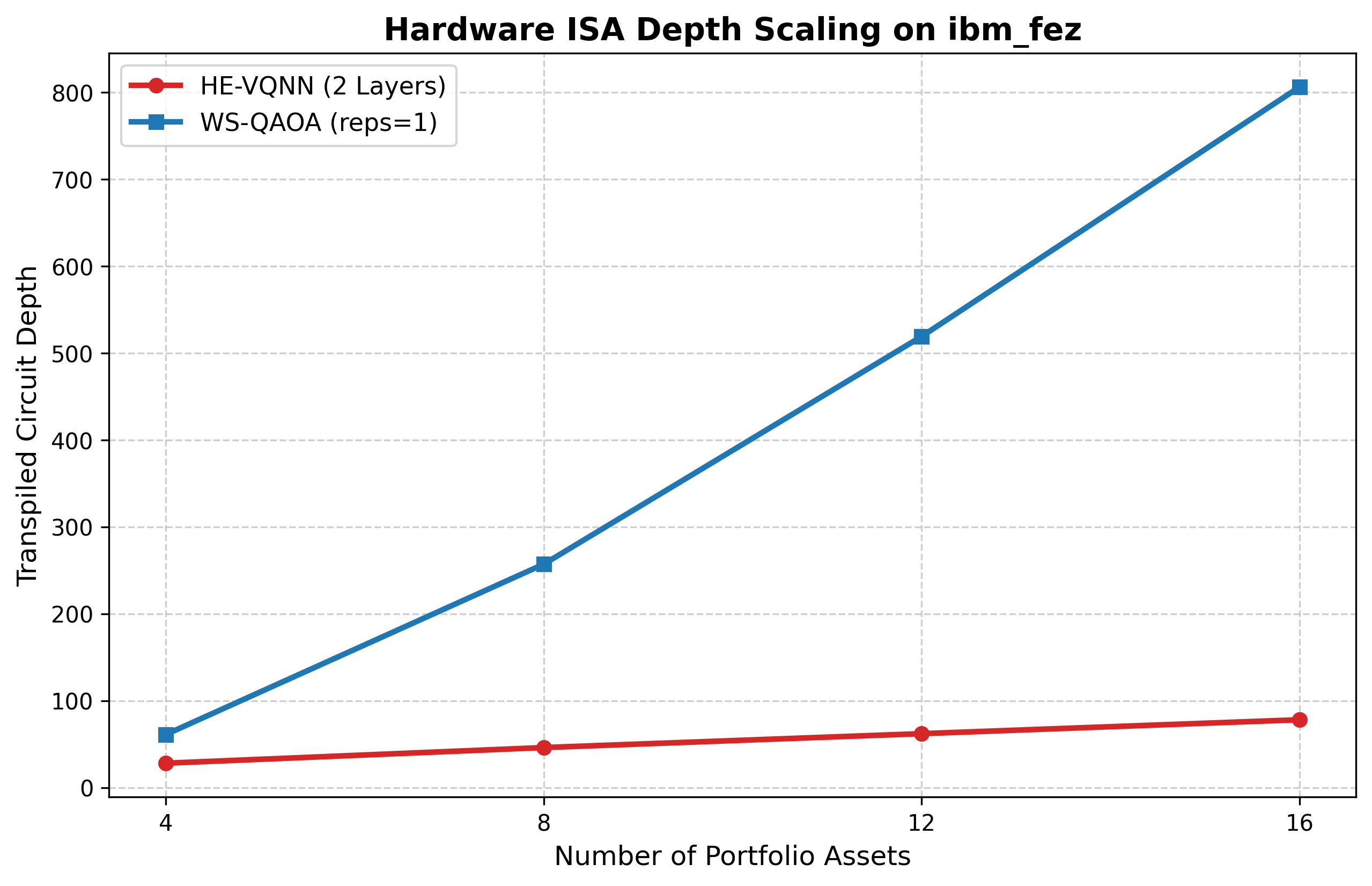}}
        \caption{Hardware ISA depth scaling.}
        \label{fig:depth_scale}
    \end{subfigure}
    \hfill
    \begin{subfigure}[h]{0.48\textwidth}
        \centering
        \fbox{\includegraphics[width=\textwidth, height = 0.65\textwidth]{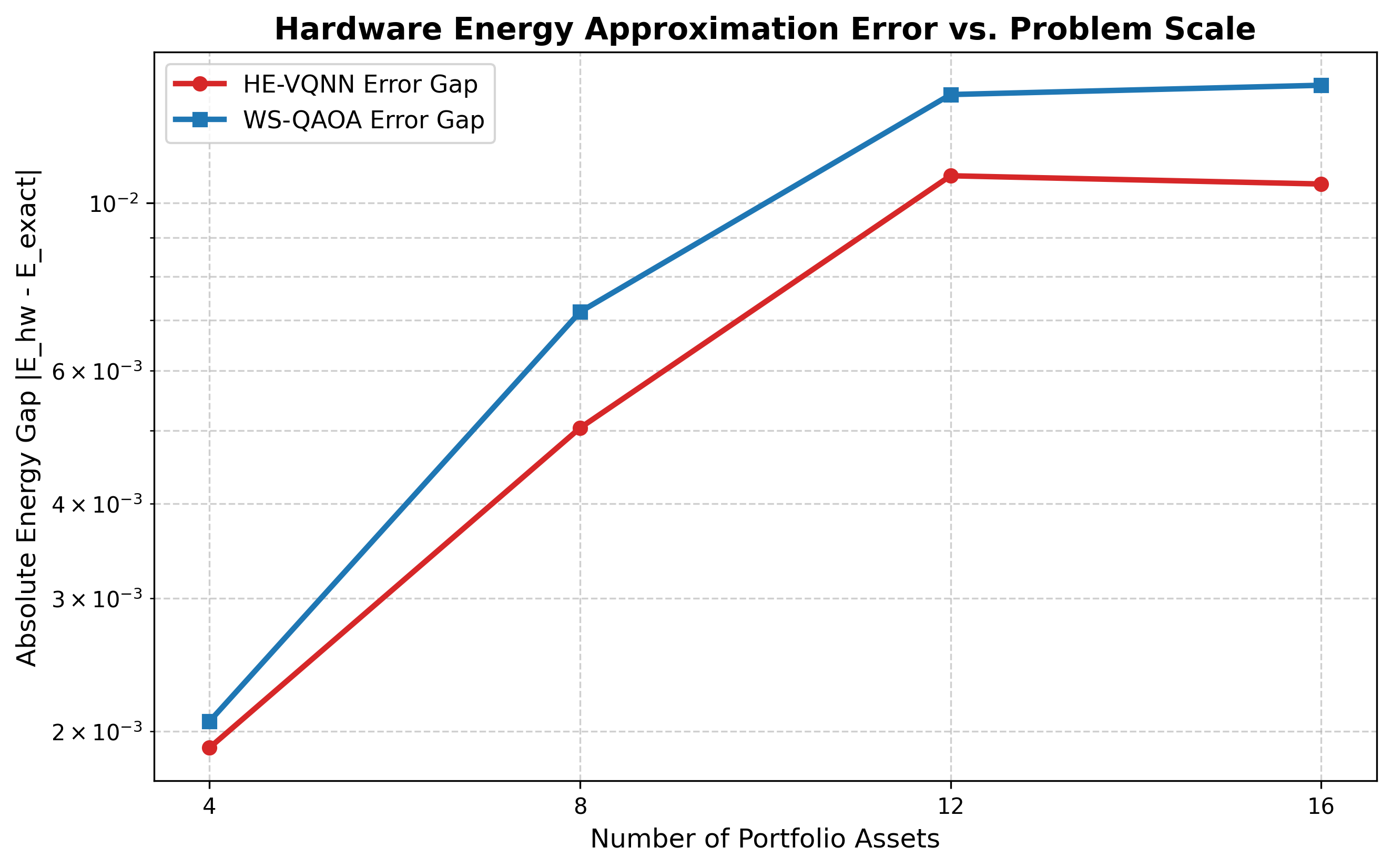}}
        \caption{Energy Approximation Error.}
        \label{fig:error_gap}
    \end{subfigure}
    \caption{Topological constraints and scaling analysis across varying asset pool sizes (4 to 16 assets). WS-QAOA suffers exponential physical overhead and decoherence growth. The hardware efficient ansatz maintains a shallow footprint, but its lack of expressibility causes a steady divergence from the optimal energy state.}
    \label{fig:scaling_analysis}
\end{figure*}

To systematically quantify these constraints, the portfolio optimization problem was scaled dynamically across $N \in \{4, 8, 12, 16\}$ assets, and the entire transpilation and execution pipeline was repeated. The collective findings are visualized in Figure \ref{fig:scaling_analysis}. As expected mathematically, Figure \ref{fig:scaling_analysis}(a) demonstrates that WS-QAOA scales exponentially in physical depth due to the heavy hex topology necessitating geometrically expanding SWAP networks to satisfy the expanding $K_N$ CVaR interactions. By contrast, the hardware aware HE-VQNN maintains a flat, stable scaling vector regardless of asset count.

This routing penalty directly and fatally dictates circuit duration. At 16 assets, the WS-QAOA circuit depth pushes the computation duration far beyond the physical $T_2$ dephasing limits of the underlying superconducting hardware, ensuring 100\% data corruption.

The ultimate consequence of this hardware level overhead is evaluated in Figure \ref{fig:scaling_analysis}(b). Plotted on a logarithmic scale, the absolute energy approximation error gap ($|E_{hw} - E_{exact}|$) is shown to grow substantially for both paradigms. This data mathematically confirms that beyond 12 assets, both algorithms suffer heavily from intrinsic limitations, establishing hard operational upper limits for dense financial systems on unmitigated NISQ devices. 

\section{Conclusion}

This study provides a rigorous, hardware based comparative benchmarking of two fundamentally distinct quantum algorithmic paradigms against highly dense Markowitz CVaR financial optimization problems. By developing and implementing a novel hybrid proxy matrix to bypass the standard CVaR auxiliary qubit bottleneck, we successfully dynamically scaled the portfolio optimization problem from 4 to 16 assets and executed the complex circuits directly on physical superconducting hardware. 

The core novelty of this research lies in explicitly identifying and quantifying the "Expressibility vs. Coherence" trap inherent to current sparse quantum architectures. The mathematical integration of CVaR provides superior asymmetric tail risk management for investors, but it natively generates dense QUBO matrices mathematically representative of complete, nonplanar graphs. 

The empirical hardware results decisively confirm that while WS-QAOA provides rigorous theoretical problem mapping, the physical manifestation of the "SWAP tax" on current planar heavy hex topologies induces severe, fatal decoherence prior to execution completion at scale. Conversely, restricting quantum entanglement strictly to physically adjacent lattice nodes allows the HE-VQNN to completely decouple its physical depth from the financial problem's density, successfully preserving quantum signal fidelity. However, this shallow, hardware efficient structure fundamentally lacks the mathematical expressibility required to capture complex, multi asset financial correlations during market downturns, resulting in broad, suboptimal probability distributions. 

Consequently, achieving genuine quantum utility for dense financial systems in the NISQ era will demand fundamentally new entanglement paradigms that intelligently balance strict hardware topological limits with the deep correlational expressibility demands of modern risk aware algorithms.

\section{Future Scope}

Future implementations of dense financial QUBOs on NISQ hardware will require novel, hybrid solutions to bridge this divide. To recover algorithmic fidelity for QAOA beyond the 12 asset limit observed in this study, the aggressive integration of advanced error mitigation frameworks, such as Zero Noise Extrapolation (ZNE) \cite{Li2017} and Probabilistic Error Cancellation (PEC) \cite{Temme2017, Endo2018}, is absolutely necessary to manage the SWAP tax. 

Conversely, to improve the HE-VQNN, researchers must focus on developing hardware aware ansatzes that offer significantly higher expressibility such as dynamically tailored entanglement graphs \cite{Cincio2021} that pulse based on specific asset correlations without triggering the exponential SWAP routing penalty. Finally, embedding targeted quantum evaluation subroutines within broader classical tensor network frameworks may allow for the macrooptimization of asset vectors exceeding 50 parameters, without directly exposing the fragile quantum hardware to the full $O(N^2)$ density of the CVaR interaction graph.

\FloatBarrier 
\balance

\end{document}